%% file: main.tex
\documentclass[conference]{IEEEtran}
\IEEEoverridecommandlockouts
\usepackage[utf8]{inputenc}
\usepackage[T1]{fontenc}
\usepackage[hyphens]{url}
\usepackage[breaklinks, colorlinks, linkcolor=black, anchorcolor=black, citecolor=black, urlcolor=black]{hyperref}
\usepackage{amsmath,graphicx,makecell}
\usepackage{enumitem} 
\graphicspath{{./fig/}}
\usepackage{threeparttable}
\usepackage{booktabs}
\usepackage{algorithmic}
\usepackage[linesnumbered, ruled, vlined]{algorithm2e}
\usepackage{adjustbox}
\usepackage{booktabs}
\usepackage{amssymb}
\usepackage{wrapfig}
\usepackage{multicol}         
\usepackage{multirow}         
\usepackage{url}
\usepackage{tabularx}
\usepackage{xurl}
\usepackage{cite}
\usepackage{amsmath,amssymb,amsfonts,amsthm}
\usepackage{xcolor}
\usepackage{soul} 
\usepackage{algorithmic}
\usepackage{graphicx}
\usepackage{textcomp}
\usepackage{url}
\usepackage{tabularx}
\usepackage{xurl}
\usepackage{threeparttable}
\usepackage{xcolor}
\usepackage{caption}
\usepackage{subcaption}
\usepackage{amsmath,amssymb}

\definecolor{midnightblue}{rgb}{0.1, 0.1, 0.44}

\usepackage{color}

\def\BibTeX{{\rm B\kern-.05em{\sc i\kern-.025em b}\kern-.08em
    T\kern-.1667em\lower.7ex\hbox{E}\kern-.125emX}}
\begin{document}

\title{Modeling and Analysis of Utilizing Cryptocurrency Mining for Demand Flexibility in Electric Energy Systems: A Synthetic Texas Grid Case Study
}
\author{\IEEEauthorblockN{Ali Menati, Kiyeob Lee, and  Le Xie }\\
 \IEEEauthorblockA{Department of Electrical and Computer Engineering\\ Texas A\&M University \\ College Station, Texas, United States \\
\text{\{menati, kiyeoblee, le.xie\}}@tamu.edu  }
 }
\maketitle

\begin{abstract}
The electricity sector is facing the dual challenge of supporting increasing level of demand electrification while substantially reducing its carbon footprint. Among electricity demands, the energy consumption of cryptocurrency mining data centers has witnessed significant growth worldwide. If well-coordinated, these data centers could be tailor-designed to aggressively absorb the increasing uncertainties of energy supply and, in turn, provide valuable grid-level services in the electricity market. In this paper, we study the impact of integrating new cryptocurrency mining loads into Texas power grid and the potential profit of utilizing demand flexibility from cryptocurrency mining facilities in the electricity market. We investigate different demand response programs available for data centers and quantify the annual profit of cryptocurrency mining units participating in these programs. We perform our simulations using a synthetic 2000 bus ERCOT grid model, along with added cryptocurrency mining loads on top of the real-world demand profiles in Texas. Our preliminary results show that depending on the size and location of these new loads, we observe different impacts on the ERCOT electricity market, where they could increase the electricity prices and incur more fluctuations in a highly non-uniform manner. 
\end{abstract}

\begin{IEEEkeywords}
cryptocurrency mining, demand flexibility, electricity market, demand response
\end{IEEEkeywords}

\input{1-Introduction}

\input{2-Demand_Response}
\input{3-Interconnection}

\input{4-Formulation}

\input{5-Synthetic_Grid}
\input{6-Simulation}

\input{7-Discussion}

\input{9-Conclusion}
\Urlmuskip=0mu plus 1mu\relax  
\bibliographystyle{IEEEtran}
\bibliography{ref}
\end{document}

%% file: 1-Introduction.tex
\section{Introduction}
With increasing electrification and integration of more renewable energy resources, power grid management is facing new challenges in terms of reliable and cost-effective operation \cite{el2022impact, white2021quantifying}. Large supply-demand mismatches, high price variations, and more frequent extreme weather events could hinder further demand electrification and renewable energy deployment. Demand flexibility plays a pivotal role in addressing these challenges. Demand flexibility improves grid reliability by closing the supply-demand mismatch gap and providing reserve during large supply scarcity events \cite{wu2022much,amadeh2022quantifying, 7478165}. It is an essential component of energy transition and reduces the need for large energy storage systems to mitigate the impact of extreme weather events like the 2021 Texas power outage \cite{menati2021preliminary}. To harness the benefits of demand flexibility, various demand response programs are designed to change the normal energy consumption patterns by shifting or reducing load in certain hours. Demand response capacities of load resources are then collected through demand response providers, which are considered "virtual generators" authorized to participate in most U.S. electricity markets~\cite{Nyiso, Caiso, Loadresource}. 

An emerging source of flexibility in modern power systems is the cryptocurrency mining capacity of data centers. Cryptocurrency mining, in particular, the bitcoin mining power capacity has nearly doubled between 2019 and 2021, and currently, its annual worldwide electricity consumption stands at around 131 TWh \cite{Bitcoinpower}. Given their substantial demand flexibility, they can help the grid achieve better load balancing through capacity right sizing and load shifting. According to Cambridge Centre for Alternative Finance \cite{Bitcoinpower}, China’s global hash rate, the computational power required to mine new Bitcoins, fell from over 75 percent to 21 percent of the worldwide total between September 2019 and January 2022. In the same period, the United States hash rate share has notably increased from nearly 4 percent to 37.8 percent, bringing up its share of power capacity to 5.7 GW as shown in Fig. \ref{fig:bitcoin}. 

A conceputually similar type of demand flexibility can also be obtained from Internet data centers. There exist a large body of work on data center workload management and scheduling to address operational cost minimization from data centers' point of view \cite{Rao12, liu2013data, Ghamkhari13, Chen16}. In \cite{liu2013data} the authors combine workload scheduling and local power generation to perform demand response by avoiding the coincident peak. 
In \cite{Cupelli18}, a control framework is presented to coordinate the cooling system, workload execution, and energy storage for enabling demand response participation. Another line of research is studying utility companies' optimal pricing mechanism to encourage data centers' participation in demand response \cite{Wang16, liu2011greening, Tran16, Bahrami19}. A comprehensive survey of the opportunities and challenges for data centers participating in demand response can be found in \cite{wierman2014opportunities}. However, compared with Internet data centers, cryptocurrency mining facilities exhibit a much higher degree of flexibility as there is little or no time sensitivity in turning on the mining facilities.

Within the United States, Texas is one of the most attractive destinations for new mining facilities. Electricity costs are a substantial part of mining facility expenditures, and the unique characteristics of the Texas electricity market managed by the Electric Reliability Council of Texas (ERCOT) provide an opportunity to acquire cheap energy through bilateral contracts. Renewable energy integration has also been growing, particularly in West Texas, where there are abundant wind energy resources. Extensive demand response programs in ERCOT are another attractive characteristic that creates a revenue stream for mining facilities with highly flexible demand. As a result, there is approximately 1.5 GW of Bitcoin mining power capacity active in Texas, which represents nearly 10 percent of the global market, and it is also attracting around 2 GW of additional Bitcoin mining capacity per year \cite{Bitcointexas}.

Rapid integration of the new Bitcoin mining loads, without proper planning and infrastructure development, creates numerous challenges for Texas electric grid. It has been shown that even a low penetration of small scale cryptocurrency miners can substantially increase the utilization of distribution transformers \cite{Moghimi21} and increase power market price volatility \cite{KARMAKAR2021110111}. These large loads might violate physical transmission line capacity constraints and impact grid reliability. Furthermore, depending on the location and the size of mining loads, they could potentially increase electricity prices in a highly non-uniform manner. Due to the critical and disproportionate effects of electricity prices on all market participants in ERCOT, studying these new loads is of great necessity. The unique attributes of the new loads, including high flexibility and fast response time, make these types of studies different from conventional loads. In this paper, we study the impact of new mining loads on the synthetic ERCOT electric grid and how their demand flexibility could help reduce grid pattern disruptions. Our main contributions in this paper are summarized as follows: 
\begin{figure}   

\begin{center}
\includegraphics[width=1\columnwidth]{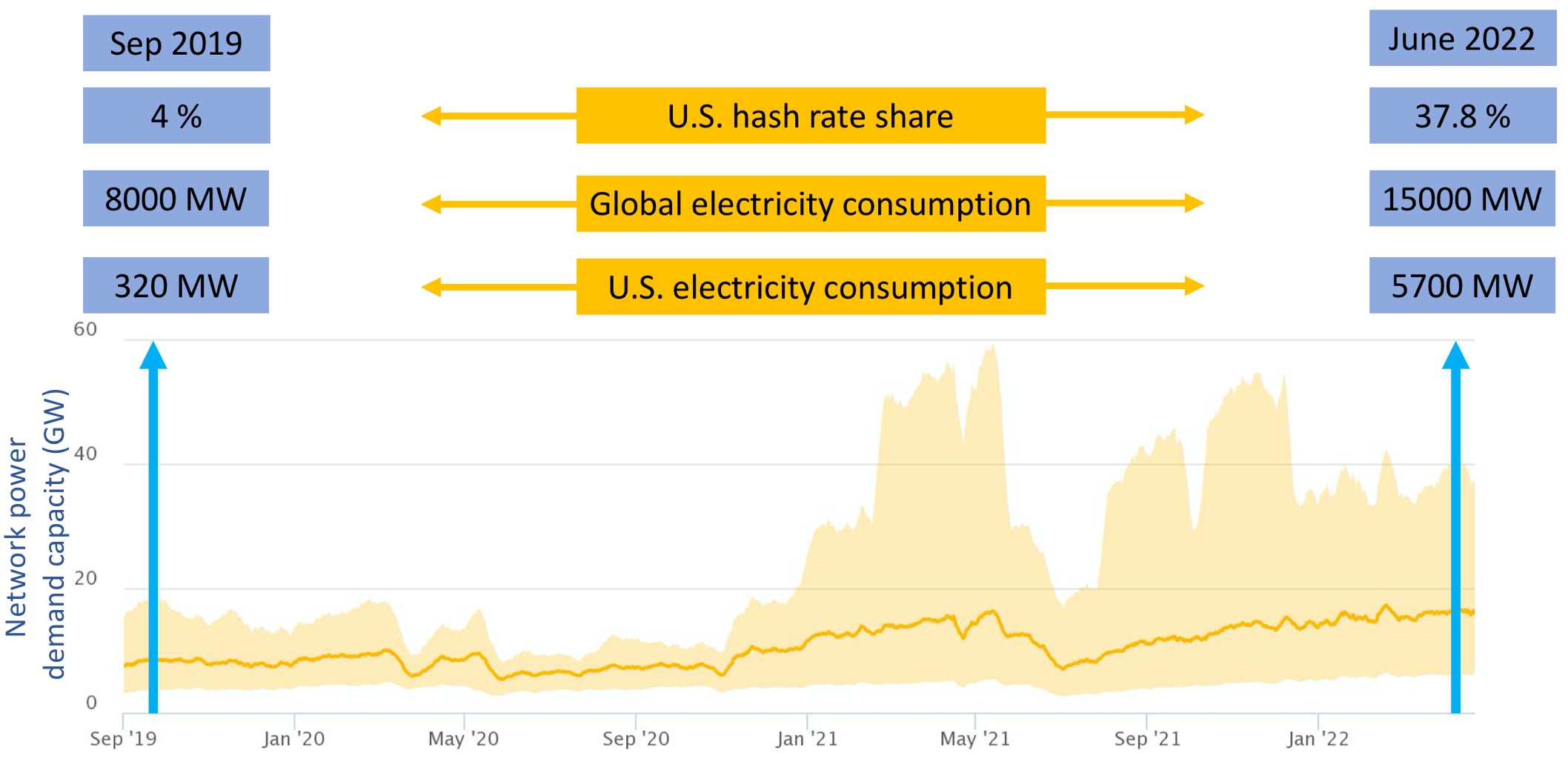}
\par\end{center}
\caption{\label{fig:bitcoin}United States share of the global Bitcoin electricity consumption and hash rate.}
 \vspace{-2mm}
\end{figure}

\begin{enumerate}

\item We introduce the role of cryptocurrency mining facilities in the ERCOT electricity market, their interaction with other stakeholders, and how they can utilize their demand flexibility by participating in the relevant demand response programs. The annual per-unit profit for one MW of demand flexibility is quantified and simulated, considering the influential parameters including Bitcoin price, electricity price, mining difficulty (amount of electricity needed for mining a single Bitcoin), and predicted demand response deployment rate. 

\item The decisive impact of new cryptocurrency loads integrated into ERCOT is studied using a 2000 bus synthetic grid model to simulate the real-time ERCOT electricity market. The Security-Constrained Unit Commitment (SCUC) and Security-Constrained Economic Dispatch (SCED) problems are solved to obtain locational marginal prices. To maximize the cryptocurrency mining data center's profit for participating in different demand response programs, we formulate their decision-making process as an optimization problem.

\item  We choose 3 locations with multiple mining facilities in the synthetic Texas grid to integrate different amounts of new cryptocurrency mining demand on top of the typical demand profile in ERCOT. The tempo-spatial properties of Locational Marginal Prices (LMPs) are analyzed by adding new mining loads with different capacities and across multiple locations. Our results show that integrating these new loads substantially increases the electricity prices during peak hours. Our county-level electricity price analysis suggests that this impact highly depends on the size and location of the new loads and the transmission line capacity constraints.

\end{enumerate}
In Section~\ref{demand_response} potential demand response programs for mining facilities are introduced. In Section~\ref{large_load} we explain the ERCOT electricity market structure, various entities interacting with cryptominers, and the importance of large load interconnection studies. The problem formulation for cryptocurrency mining facility participation in demand response programs is presented in Section~\ref{problem}. Section~\ref{synthetic} introduces the synthetic 2000 bus grid model used in our study and the experiment set-up for running the simulation. The results and findings of our simulations are presented in Section~\ref{simulation}. Section~\ref{sec:Discussion} discusses the importance of our findings and the possible implications of these results on long-term policy planning. Finally, Section~\ref{sec:conc} concludes the paper.

%% file: 2-Demand_Response.tex
\section{Demand response programs}
\label{demand_response}
\begin{table*}[htb]
    \centering
    \caption{Characteristic of demand response programs in ERCOT \cite{2021annual, Potomac, ercot_dec}}
    \begin{tabular}{lccc} \toprule
         & \begin{tabular}[c]{@{}l@{}}Average Price (\$/MWh)
\end{tabular} 
         & \begin{tabular}[c]{@{}l@{}}Total Capacity Procured (MW)\end{tabular} 
         & \begin{tabular}[c]{@{}l@{}}Procurement Time-Scale
\end{tabular}    \\ \midrule 
        RRS & 11.27  & 1564 & Day-ahead
 \\
        ERS  & 6.03 & 925  &  4 months
 \\
        Price-driven & program specific  & 2800 & program specific\\
        \bottomrule
    \end{tabular}
    \label{tab:DR_parameters}
\end{table*}

In ERCOT, Bitcoin mining loads are considered Controllable Load Resources (CLRs), which are eligible to perform all ancillary services and participate in the real-time energy market \cite{drercot}. Here we present the most prominent demand response programs available for mining facilities in Texas.

\subsection{Ancillary Service Market} 
In ERCOT electricity market, load resources are authorized to participate in the ancillary service market to perform Responsive Reserve Service (RRS), non-spinning reserve service, and regulation up/down. Among these services, the largest one is RRS, where the load is automatically cut off by under-frequency relays or manually interrupted within a ten-minute notice in case of need for load reduction~\cite{habib2021improving}. RRS is proven to be very impactful, and during the Texas power outage of February 2021, load resources providing RRS reduced their consumption by more than 1,400 MW~\cite{king2021timeline}.
    
\subsection{Emergency Response Service (ERS)}
While RRS is activated during large supply scarcity events, there is another demand response program in ERCOT called Emergency Response Service (ERS), which is only deployed as the last reserve to avoid load shedding during grid emergencies. In addition, ERS is not an ancillary service, and it is procured by ERCOT four times a year in two different response times of 10 and 30 minutes~\cite{drercot}. On average, ERCOT keeps around 1,000 MW of ERS resources as emergency reserve~\cite{drercot}. 

\subsection{Price-Driven Demand Response} 
In addition to providing reserve in electric energy systems, incentive-based and price-responsive load management programs are designed to exploit the elasticity of industrial and residential loads during high electricity price hours~\cite{zhong2012coupon,xia2017energycoupon,ming2020prediction}. These programs are mostly administered by Transmission and Distribution Service Providers (TDSPs). An essential goal of
these programs is to keep the electricity prices in the system as
low as possible. For instance, in the 4-Coincident Peak (4CP) program in ERCOT, the single monthly 15-minute system peak during each of the four months from June through September is set, and the 4CP charge for distribution service providers is determined by calculating their load during ERCOT's monthly demand peak. To avoid these substantial charges, load entities seek to reduce their demand by offering incentives and rewards depending on the system condition. According to ERCOT, the 4CP load reduction capacity during peak hours is up to 2,800 MW \cite{Potomac}. The most prominent aspects of demand response programs are compare in Table \ref{tab:DR_parameters}.

%% file: 3-Interconnection.tex
\section{ERCOT Electricity Market Structure}
\label{large_load}

\begin{figure}   

\begin{center}
\includegraphics[width=\columnwidth]{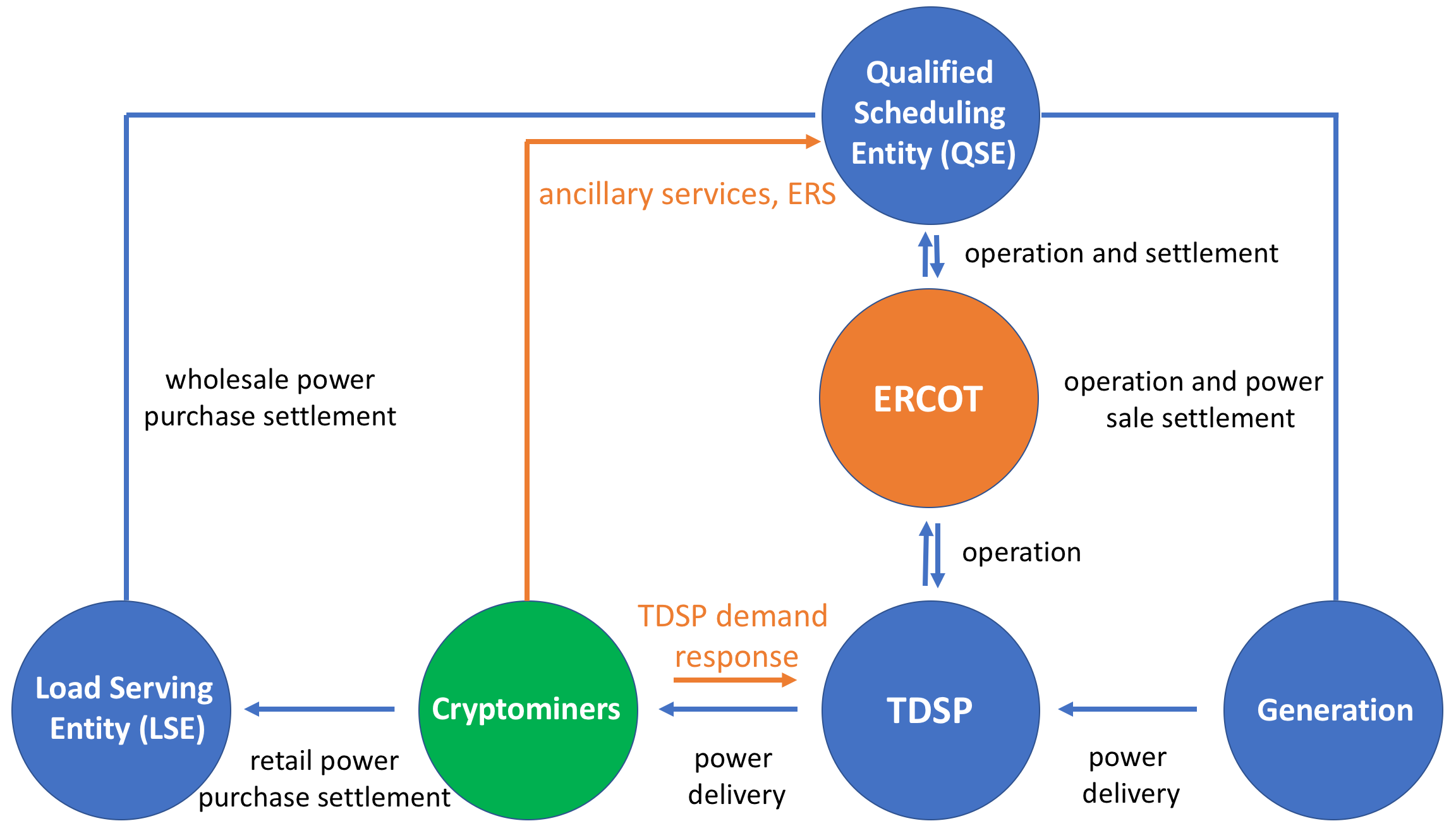}
\par\end{center}
\caption{\label{fig:ercot}ERCOT electricity market structure and the entities interacting with cryptocurrency mining facilities.}
\end{figure}

\subsection{Market Participants}
There are various entities with different objectives interacting in ERCOT electricity market. In the following subsections, we present the key entities influencing mining facilities. 

\begin{itemize}
    \item \textbf{Qualified Scheduling Entities (QSEs)} are authorized to submit bids and offers for resource entities to sell energy or on behalf of load serving entities to purchase energy in the day-ahead and real-time markets. As shown in Fig. \ref{fig:ercot} QSEs are also qualified to represent cryptocurrency miners and participate in the ancillary service market or submit ERS bids on their behalf. If an offer is accepted by ERCOT, the QSE gets paid according to the settlement price. In addition, QSE is responsible for mining facility modeling, telemetry, and possible outage scheduling.

    \item \textbf{Load Serving Entities (LSEs)} are authorized to represent competitive retailers selling electricity to retail customers. Load forecasting and negotiating with power producers to buy cheap energy is done by LSEs, and they participate in the wholesale market through a QSE, as shown in Fig. \ref{fig:ercot}. It can also be seen that mining facilities usually acquire their electricity services through an LSE, and their electricity gets delivered by Transmission and Distribution Service Providers (TDSPs). 
    \item \textbf{Cryptocurrency Miners} are capable of participating in various demand response programs. As demonstrated in Fig. \ref{fig:ercot}, some of these programs are administered by TDSPs, which handle operation scheduling and profit management. On the other hand, if the mining facility wants to participate in the ancillary service market or provide ERS, their corresponding QSE submits their offer and communicates their possible operational details. 
    \item \textbf{ERCOT} is in the center of the electricity market operation in Texas, and it handles both physical supply-demand balancing management and the market clearing process. As it can be seen from Fig. \ref{fig:ercot}, QSEs submit electricity sale offers of generation units and power purchase bids of the LSEs to ERCOT, and after performing the market clearing process, ERCOT returns the settlements along with the operational control signals of the generation units through QSEs. At the same time, ERCOT communicates operational control signals with TDSPs to ensure safe and economic grid operation. 
    
\end{itemize}

\begin{figure}   

\begin{center}
\includegraphics[width=.95\columnwidth]{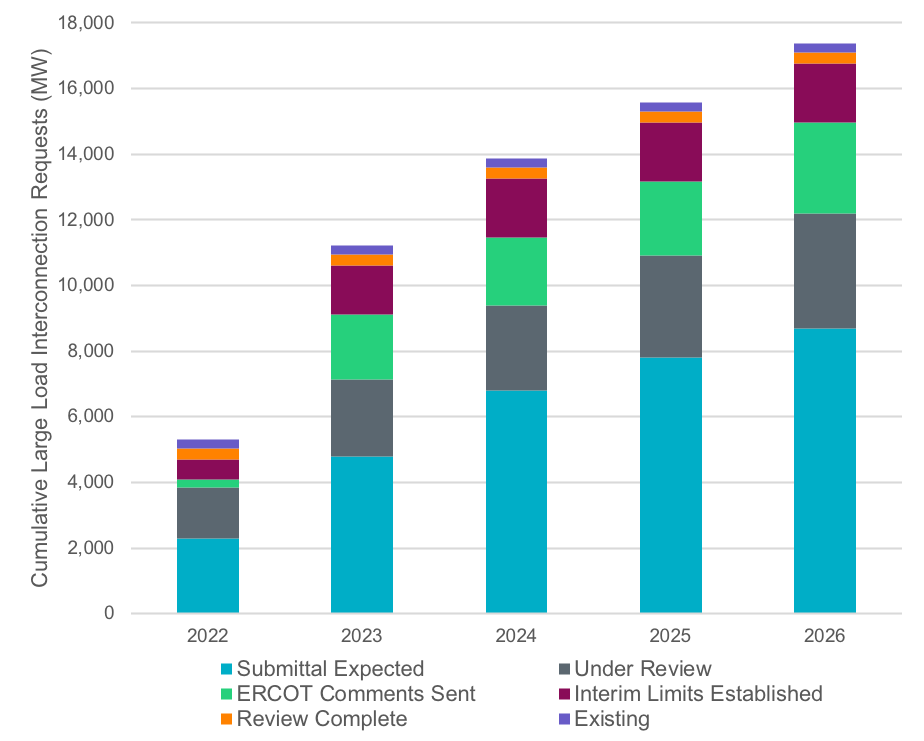}
\par\end{center}
\caption{\label{fig:largeload}Large load interconnection request projection in ERCOT between 2022 and 2026 \cite{largeload}.}
\vspace{-2mm}
\end{figure}
\subsection{Large Flexible Load Interconnection in ERCOT}
The rate of cryptocurrency mining data center integration into ERCOT grid is faster than standard transmission network expansion plans. In addition, due to their large flexibility, voltage control of these loads and their disruptive impact on the electricity market are vastly different from conventional static loads. Hence, it is imperative for the miners, policy-makers, and grid operators to study their interconnections and plan accordingly. Depending on the location and the size of the new mining facilities being built in Texas, we observe substantially different impacts on the grid. If they are relatively small they have minor impacts, but if they are larger, they increase transmission congestion and electricity prices, and finally if they are very large they push the system solution into infeasibility. In our case study, we observed all three cases when installing different load sizes across the synthetic Texas grid. To ensure safe integration of large cryptocurrency mining loads into the system, transmission service providers are required to perform interconnection studies for new loads not co-located with a resource with total demand of 75 MW or greater or co-located with a resource with total demand of 20 MW or greater \cite{Interim}. As shown in Fig. \ref{fig:largeload} the request for large load interconnections is growing fast, and it is expected that between 2022 and 2026, ERCOT will receive requests to add more than 17 GW of these new large loads.

%% file: 4-Formulation.tex
\section{Problem Formulation}
\label{problem}
With multiple demand response programs available for cryptocurrency mining data centers, they seek to find the optimal combination of these programs that maximizes their profit. This section formulates cryptominers demand response portfolio selection as an optimization problem. Consider a scenario in which we have a data center with a mining capacity of C (MW) at any given time. The market is constantly changing, and the cryptominer observes a different electricity price, cryptocurrency price, and demand response profit at each time slot $t$, where the set of time slots is denoted by $\mathcal{T}:=\{1, ..., T\}$. Let us denote $\mathcal{N}:=\{1, ..., N\}$ as the set of all possible demand response programs available for the cryptominer, and $c_i \leq C$ as the demand response capacity obligation of the data center under program $i$, where $i \in \mathcal{N}$. We denote $\hat{p_i}(t)$ as the expected per-unit revenue of participating in the $i$th demand response program at time $t$, and $\hat{p}(t):= (\hat{p_1}(t), \hat{p_2}(t), ..., \hat{p_N}(t))$ captures this value for all demand response programs. Depending on the program structure, the real-time participation revenue is calculated in different ways. If the cryptominer participates in ERCOT ancillary service market, $\hat{p_i}(t)$ is the market clearing price, and if it has a long-term contract like ERS, $\hat{p_i}(t)$ is calculated according to agreed upon contract specification.

We denote $\hat{d_i}(t)$ as the expected deployment rate of the $i$th demand response program at time $t$, and $\hat{d}(t):= (\hat{d_1}(t), \hat{d_2}(t), ..., \hat{d_N}(t))$ as the expected deployment rate profile of all demand response programs, where $0 \leq \hat{d_i}(t) \leq 1$. This value captures the fraction of the committed capacity that is being deployed at each moment. Hence, $c_i \hat{d_i}(t)$ is the expected deployed capacity under the $i$th demand response program at time $t$. It should be noted that demand response deployment rate is highly uncertain, and depending on the type of program, deployment rate follows different distributions, but here we assume that we are able to obtain its expected value.

Finally we define $\hat{r}(t)$ as the expected per unit net reward of mining cryptocurrency at time $t$ for one unit of electricity. This reward is calculated as $\hat{r}(t)=\hat{p_b}(t)-\hat{p_e}(t)$, where $\hat{p_b}(t)$ is the expected revenue obtained from selling the cryptocurrency mined using one unit of electricity, and $\hat{p_e}(t)$ is the expected per-unit electricity cost. For example, if at time $t$ it takes $143 MWh$ to mine a Bitcoin and the price for one Bitcoin is $\$25,000$, we calculate $\hat{p_b}(t)$ as $\frac{25000\$/BTC}{143 MWh/BTC} \approx 175 \$/MWh$. Now we proceed by introducing our 
cryptominer optimal demand response portfolio selection problem as follows:
\begin{subequations} 
\label{eq:dr} 
	\begin{eqnarray}
	&\underset{[c_i]_{i=1}^N}{\max}&  \sum_{i=1}^{N} \sum_{t=1}^{T} \big[c_i \hat{p_i}(t) -  c_i \hat{d_i}(t) \hat{r}(t) \big]\\
	&\mbox{s.t.}\;& c_i \geq 0,  \, \, \, \textit{and}  \, \, \, \sum_{i=1}^{N} c_i \leq C, \, \, \, \,  \textit{for} \, \, \, \,  i \in \mathcal{N},  \label{1d}
	\end{eqnarray} 
\end{subequations}
where $c_i \hat{p_i}(t)$, and $c_i \hat{d_i}(t) \hat{r}(t)$ are the expected revenue and loss of participating in the program i, respectively. In this formulation $\sum_{t=1}^{T} c_i \hat{d_i}(t) \hat{r}(t)$ captures the total loss of revenue due to not mining Bitcoin during deployment in the $i$th demand response program. Hence when this value is large, it is not beneficial to participate in the demand response program. Constraint~\eqref{1d} assures that the sum of participation capacity in all demand response programs does not exceed the mining capacity of the data center. As shown above, our objective is to maximize cryptominer's expected profit by choosing the optimal combination of different programs, where decision variable $(c_1, c_2, ..., c_N)$ is the demand response capacity profile of the cryptominer. It should be noted that this decision is highly related to how many times and how much the operator deploys the cryptominer in each demand response program. In our setting, we assume that the demand response participation profile stays the same during each interval $[1, T]$. However, our formulation could be easily extended to capture a more general setting, where the capacity of each program $c_i$ changes over time. This turns our problem formulation into an online optimization that is capable of fully capturing the uncertainty involved in the decision-making process. Solving the online optimization version is out of the scope of this work, so we assume that an expected value of the future parameters is obtained through prediction and historical records.

It is worth noting that the optimization problem \eqref{eq:dr} is in the form of linear programming, and by applying the principles of the simplex method, the inequality constraints define a polygonal region, where the solution is at one of the vertices. Hence, the optimal solution in this setting is either not to participate in any demand response program or participate with full capacity $C$ in the one with the maximum expected profit. This optimization problem is solved using the predicted information in the time horizon $[1, T]$. However, in our simulations, instead of utilizing predicted future information, we use a synthetic grid model and simulate the electricity market to obtain the actual value of all the information needed to quantify the profit of various demand response programs and compare their performance under different scenarios.

%% file: 5-Synthetic_Grid.tex
\section{Synthetic Grid and Experiment Set-up}\label{synthetic}

To observe the impact of adding new mining loads and possible demand response programs, we use a large-scale open-source synthetic grid that has been developed and calibrated to match the actual Texas grid \cite{birchfield2017grid, wu2021open, lee2022targeted}. This is one of the largest publicly available synthetic grid models (2000-bus) equipped with various features. Researchers have been adding public open-access data sets such as generator capacities, wind and solar generation profiles, and hourly demand profiles to this model, in order to help statistically match the real grid \cite{xu2020us}. In this project, we are using the electricity market model built on top of the synthetic grid model in \cite{lee2022targeted}. This model has been shown to provide a realistic representation of electricity market behavior and the optimal electricity prices across Texas.

\begin{figure}
\centering
\begin{minipage}[t]{.95\linewidth}
\centering
\includegraphics[width=1\linewidth]{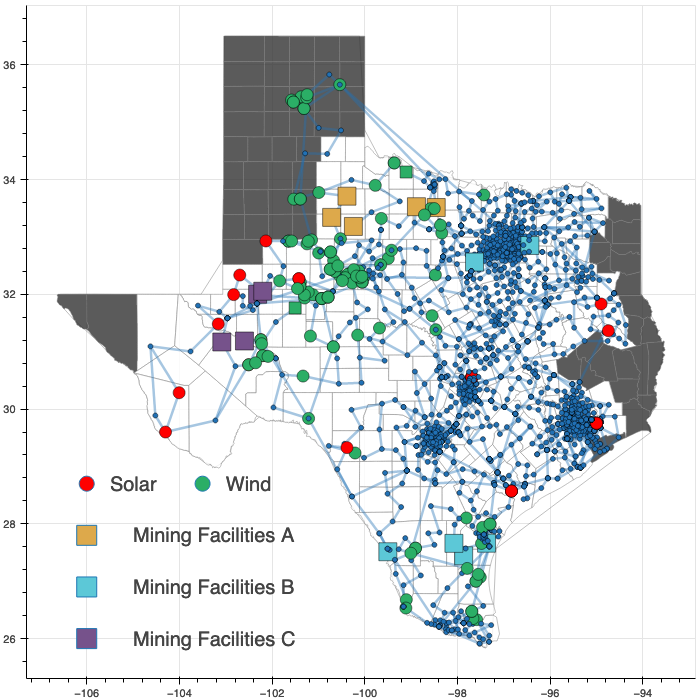}
\caption{The topology of Texas synthetic grid.}
\label{fig:texas_grid}
\end{minipage}\hfill
\vspace{-2mm}
\end{figure}

The topology of the Texas 2000-bus synthetic grid is shown in Fig. \ref{fig:texas_grid}. Observe that many buses (nodes) are clustered around the centers of population, and there are 254 counties in Texas, among which 63 counties are not monitored by ERCOT during the period considered in this study and are colored in black. Locations of renewable generators are highlighted in Fig. \ref{fig:texas_grid} and note that most renewable generators are located in western regions due to rich availability of renewable energy sources. Due to the rich source of renewable energy sources, it is also advised to locate mining facilities in western regions. The load and generation profiles for the year 2020 are obtained to implement our simulation.

In power grids, a security-constrained unit commitment (SCUC) problem is solved using day-ahead predictions to determine which units are supposed to be committed for the next day. In large-scale power systems, because of the generation units with slow ramping rates, it is necessary to solve the SCUC with bulk estimate predictions in the day-ahead market. After determining active generators, a security-constrained economic dispatch (SCED) problem is solved in real-time with more accurate predictions. SCED problems provide a more precise generation planning and determine the real-time locational marginal prices at every node. By solving SCUC and SCED problems in the synthetic grid model, wholesale and retail electricity markets are formulated, and LMPs are obtained over a period. The LMP represents the electricity price (\$/MWh) at a particular location (bus), which is defined as the extra cost incurred to serve an additional unit of load at that location. There is widely available software such as MatPower \cite{zimmerman2010matpower}, a Matlab package for solving power flow and optimal flow problems, and MatPower Optimal Scheduling Tool (MOST) for optimally scheduling generating units. MatPower and MOST are used to solve SCUC and SCED problems in the simulation. Note that SCUC and SCED problems require load profiles and renewable generation profiles (or estimates of them), and they build foundations of wholesale and retail electricity markets. In practice, there are additional layers and legacies for reliable and efficient operations of physical systems and electricity markets, however, we take a simple form of electricity markets.

%% file: 6-Simulation.tex
\section{Simulation Results} \label{simulation}

Both location and capacity of mining facilities play decisive roles and clearly have disruptive impacts when integrated into transmission grids. We investigate the disruptions of mining facilities in the synthetic grid, and potential roles and benefits of demand response for mining facilities in electricity markets. In this case study, a summer period of 61 days (Days 180 - 240) is considered since the summer period has the highest demand during a year and is often considered the most interesting time period to study. 

\subsection{Disruptions by Mining Loads}
SCUC problems are solved in a default setting without additional mining loads as a baseline for comparison. SCUC problems are also solved with additional mining loads, especially emphasizing two characteristics of mining facilities relevant to transmission grids (namely locations and capacities) as follows. Three sets of locations for mining facilities are considered, as highlighted in Fig. \ref{fig:texas_grid}. Locations of mining facilities are arbitrarily chosen yet in close proximity to the locations of renewable generators in the grid. Locations can support different amounts of additional mining loads, i.e., locations have different hosting capacity limits. For example, some locations may be able to safely host up to additional 600MW without a major upgrade to the system, while other locations may only be able to host up to additional 100MW. Thus, different total capacities of mining loads are tested out together with three sets of locations to understand disruptions induced by mining facilities.

Moreover, we assume that mining facilities are consuming a fixed amount of loads for all time steps. For example, suppose there are six mining facilities, and they consume a total of 600MW in the study period (Days 180 - 240). Then, each mining facility uniformly and equally consumes 100MW over 61 days. Given both location and capacity of mining facilities determined, SCUC and SCED problems are solved to choose online/offline status of generating units and LMPs. There are three locations (Locations A, B, C) for mining facilities in Fig. \ref{fig:texas_grid}. In Location A, there are six mining facilities and total capacities 360MW, 480MW, 600MW, 720MW, 840MW are tested (each mining facility equally consumes 60MW, 80MW, 100MW, 120MW, 140MW, respectively); In Location B, there are six mining facilities and total capacities 600MW, 720MW, 840MW, 960MW are tested; In Location C, three values of total capacity 50MW, 100MW, 150MW are tested; Lastly, a combination of Location A and B is considered that there are 12 mining facilities and total capacities 600MW, 720MW, 840MW, $\ldots$ , 1440MW are tested (each mining facility consumes 50MW, 60MW, 70MW, $\ldots $ 120MW respectively).

It is rightfully possible that mining facilities apply more sophisticated load control strategies (or policies). For example, load control strategies may be temporally and spatially reasoned that they consume more during non-peak hours and consume less during peak hours, consume more in a particular region and consume less in another region. However, we restrict to a simple load control strategy that facilities consume a fixed uniform amount of loads for simplicity.

\begin{figure*}
    \centering
    \begin{subfigure}[b]{0.475\textwidth}
        \centering
        \includegraphics[width=\textwidth]{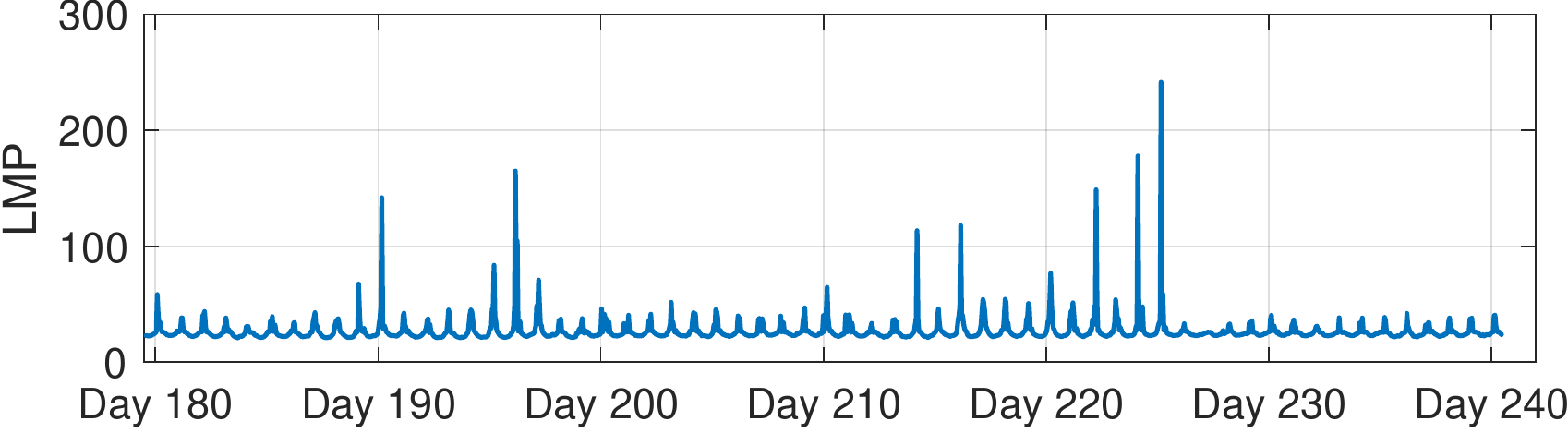}
        \caption[]%
        {{\small Baseline (without mining loads)}}
        \label{fig:original_LMP}
    \end{subfigure}
    \hfill
    \begin{subfigure}[b]{0.475\textwidth}  
        \centering 
        \includegraphics[width=\textwidth]{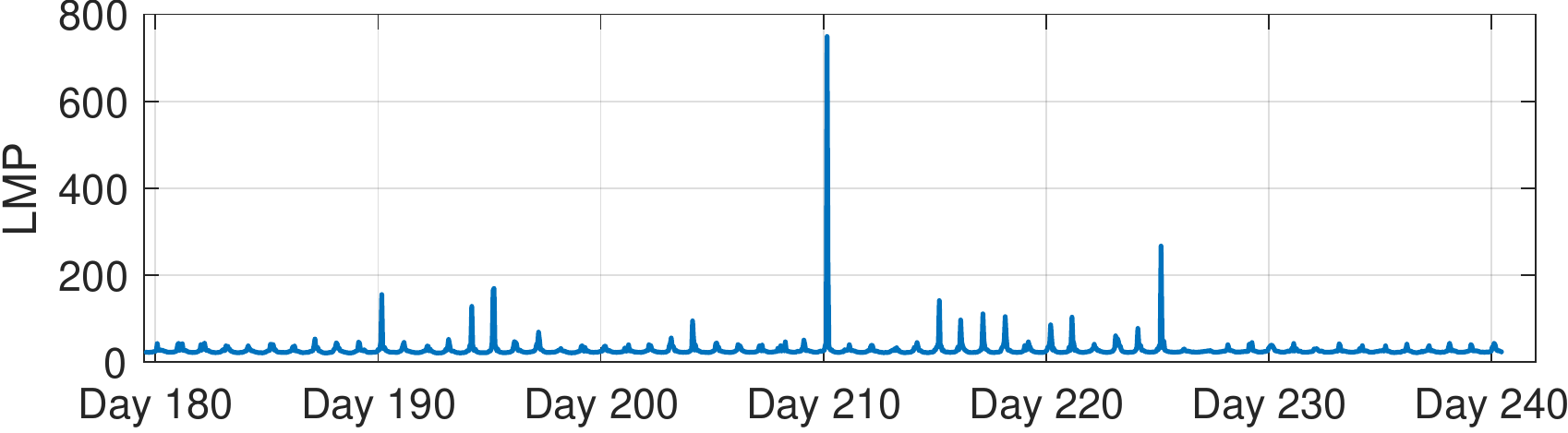}
        \caption[]%
        {{\small 600MW mining loads in Location A}}
        \label{fig:LMP_locA_600}
    \end{subfigure}
    \vskip\baselineskip
    \vspace{-1em}
    \begin{subfigure}[b]{0.475\textwidth}   
        \centering 
        \includegraphics[width=\textwidth]{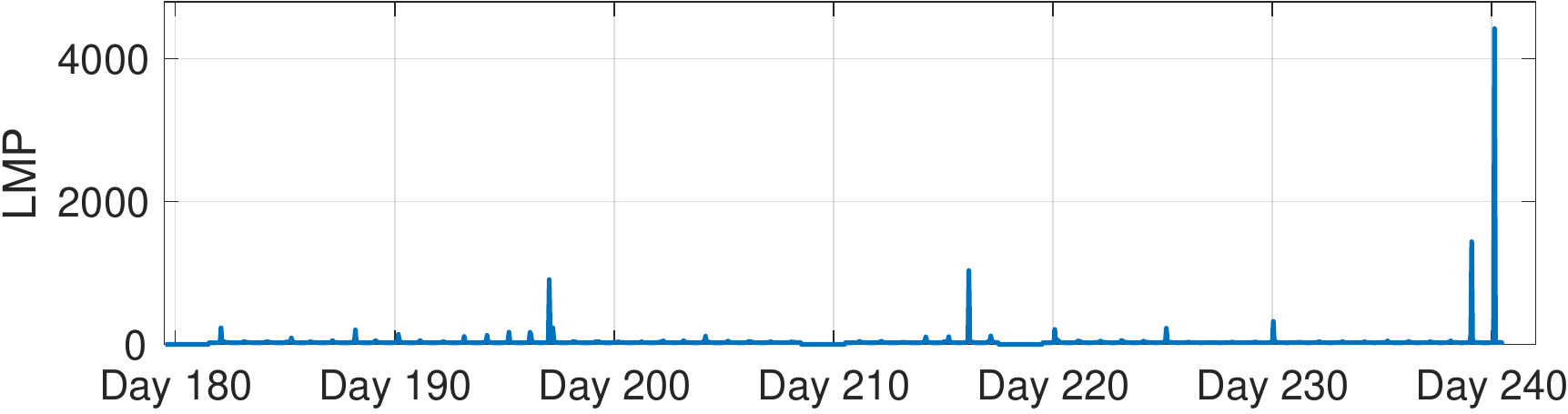}
        \caption[]%
        {{\small 720MW mining loads in Location A}}
        \label{fig:LMP_locA_720}
    \end{subfigure}
    \hfill
    \begin{subfigure}[b]{0.475\textwidth}   
        \centering 
        \includegraphics[width=\textwidth]{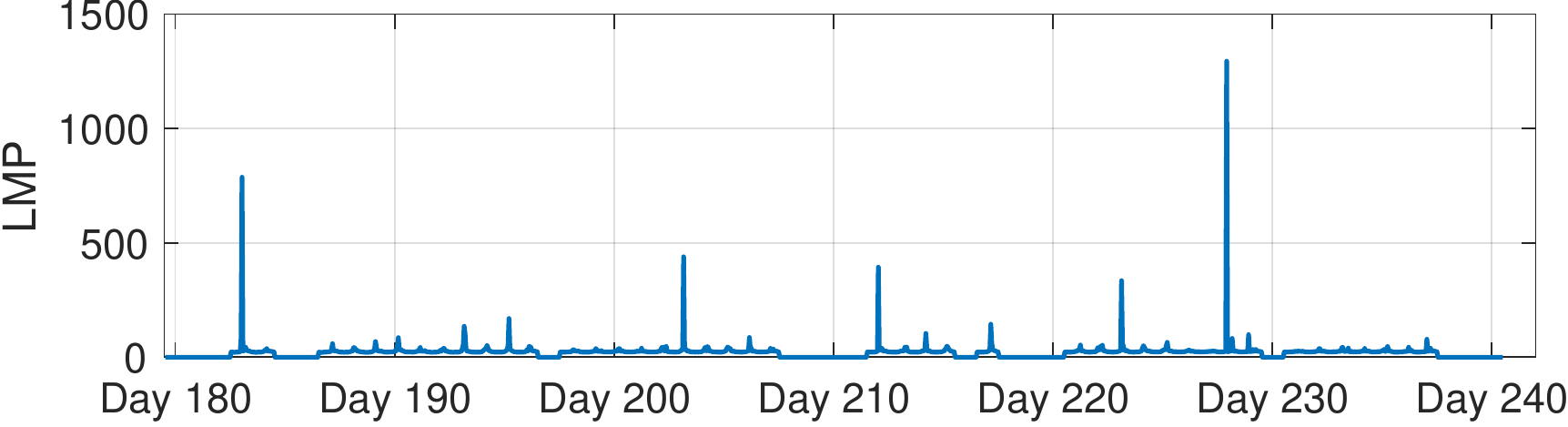}
        \caption[]%
        {{\small 840MW mining loads in Location A}}
        \label{fig:LMP_locA_840}
    \end{subfigure}
    \vskip\baselineskip
    \vspace{-1em}
    \begin{subfigure}[b]{0.475\textwidth}   
        \centering 
        \includegraphics[width=\textwidth]{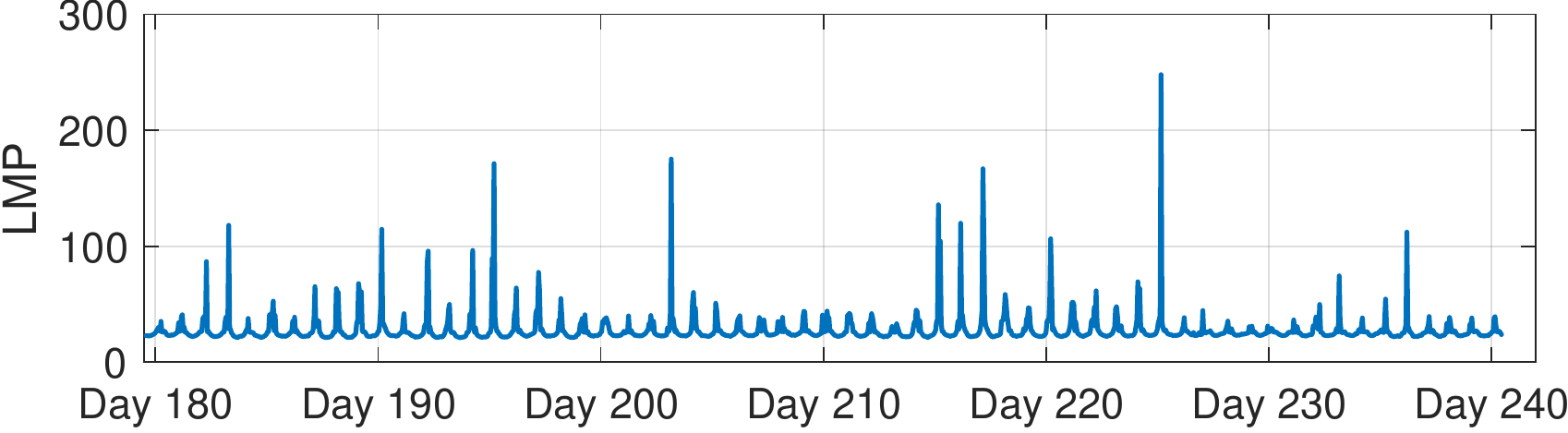}
        \caption[]%
        {{\small 840MW mining loads in Location B}}
        \label{fig:LMP_locB_840}
    \end{subfigure}
    \hfill
    \begin{subfigure}[b]{0.475\textwidth}
        \centering
        \includegraphics[width=\textwidth]{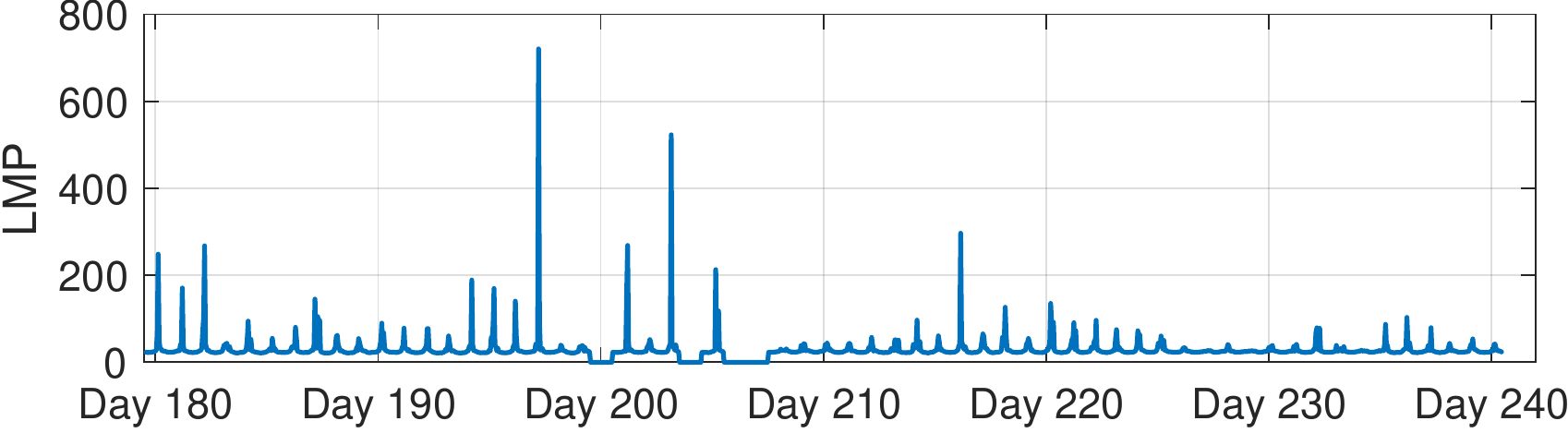}
        \caption[]%
        {{\small 960MW mining loads in Location B}}
        \label{fig:LMP_locB_960}
    \end{subfigure}
    \caption[]
    {\small The average LMP (\$/MWh) with changing locations and capacities of mining facilities in Texas synthetic grid.}
    \label{fig:avg_lmp}
\end{figure*}
\begin{figure*}
    \centering
    \begin{subfigure}[b]{0.32\textwidth}
        \centering
        \includegraphics[width=\textwidth]{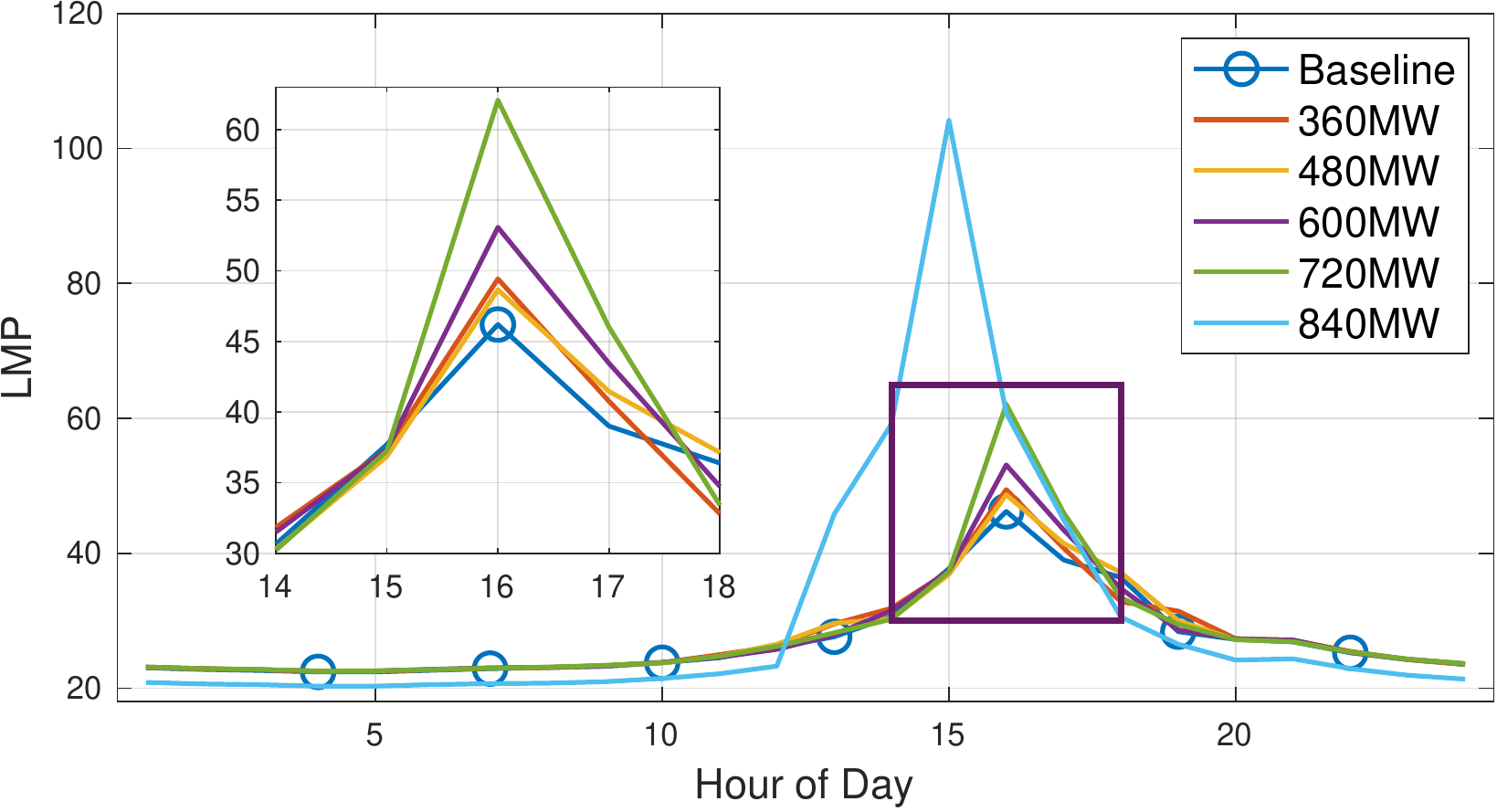}
        \caption[]%
        {{\small Location A}}
        \label{fig:hourly_lmp_locA}
    \end{subfigure}
    \hfill
    \begin{subfigure}[b]{0.32\textwidth}  
        \centering
        \includegraphics[width=\textwidth]{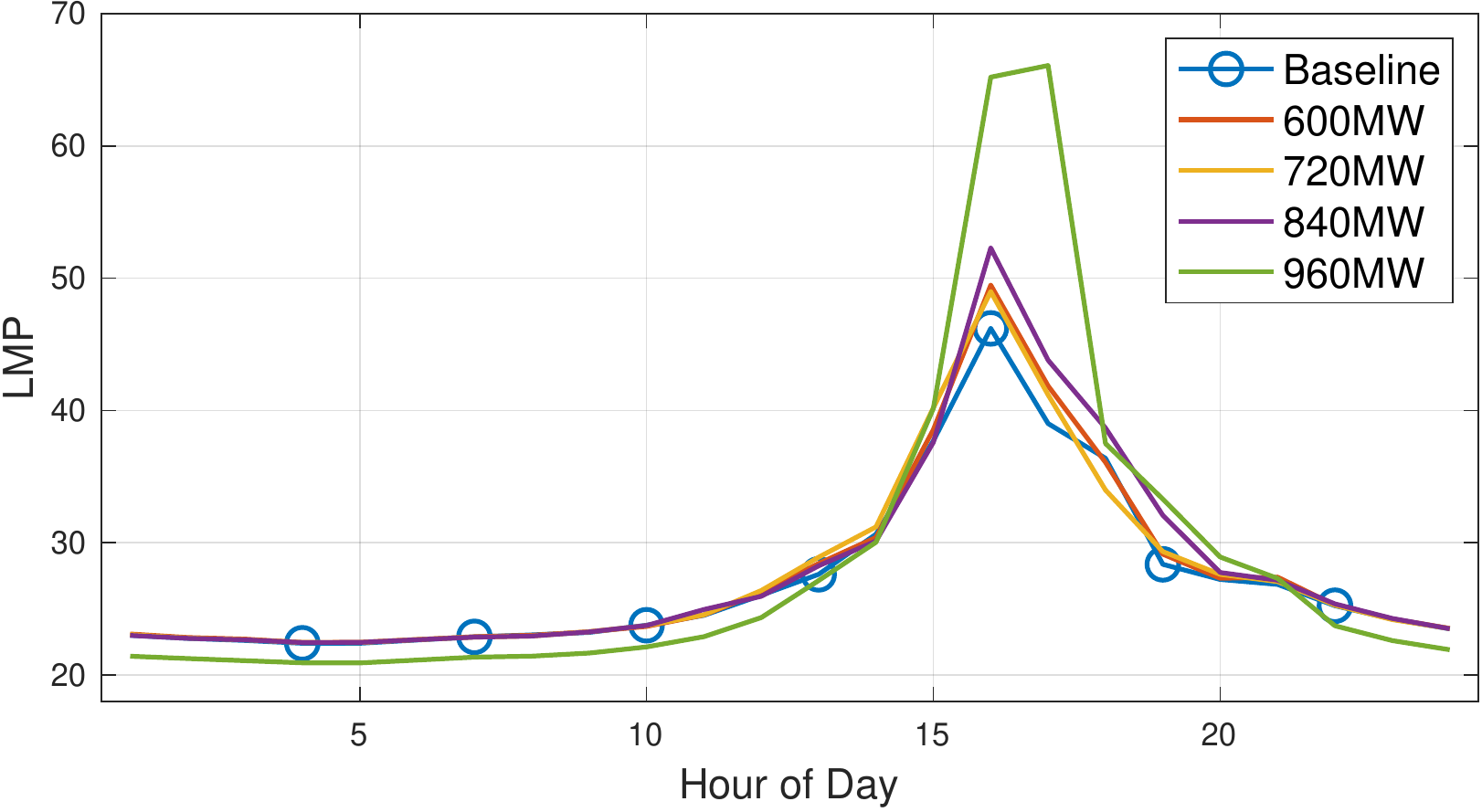}
        \caption[]%
        {{\small Location B}}
        \label{fig:hourly_lmp_locB}
    \end{subfigure}
    \hfill
    \begin{subfigure}[b]{0.32\textwidth}  
        \centering
        \includegraphics[width=\textwidth]{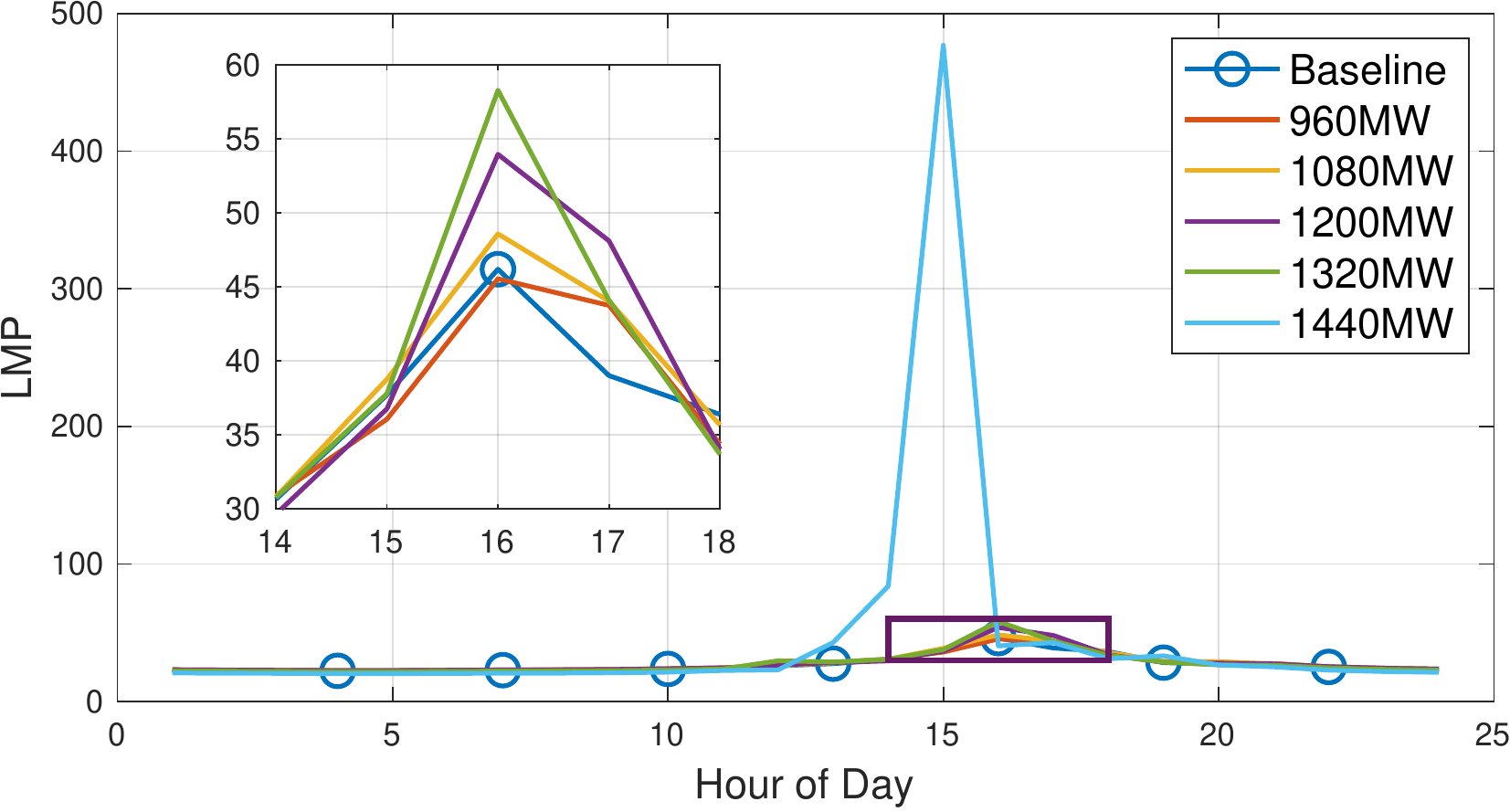}
        \caption[]%
        {{\small Location A and B}}
        \label{fig:hourly_lmp_locAB}
    \end{subfigure}
    \caption[]
    {\small Hourly LMP (\$/MWh) with changing locations and capacities of mining facilities in Texas synthetic grid.}
    \label{fig:hourly_lmp}
\end{figure*}
\begin{figure*}
    \centering
    \begin{subfigure}[b]{0.49\textwidth}
        \centering
        \includegraphics[width=\textwidth]{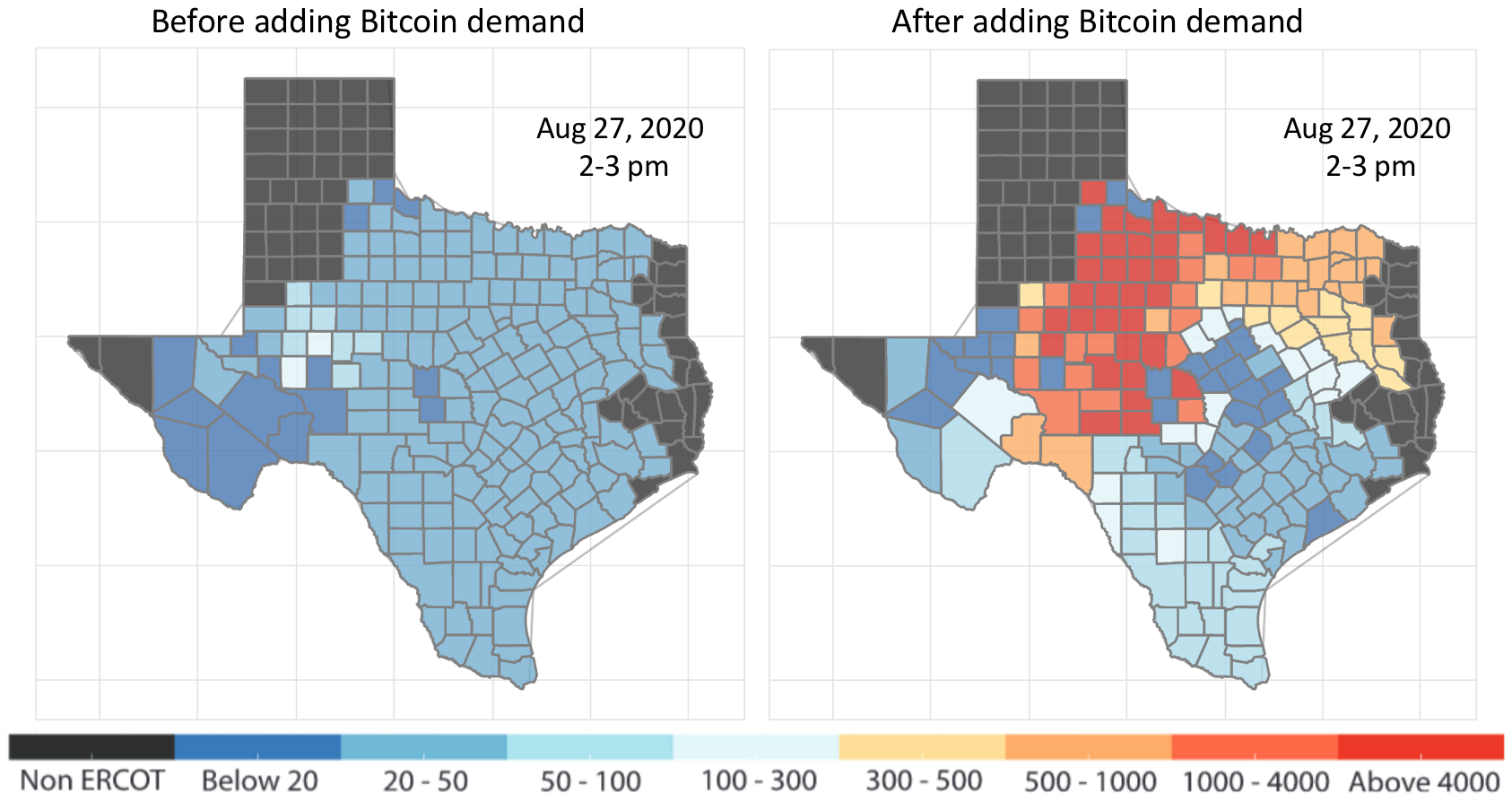}
        \caption[]%
        {{\small 720MW mining loads in Location A}}
        \label{fig:map1}
    \end{subfigure}
    \hfill
    \begin{subfigure}[b]{0.49\textwidth}  
        \centering 
        \includegraphics[width=\textwidth]{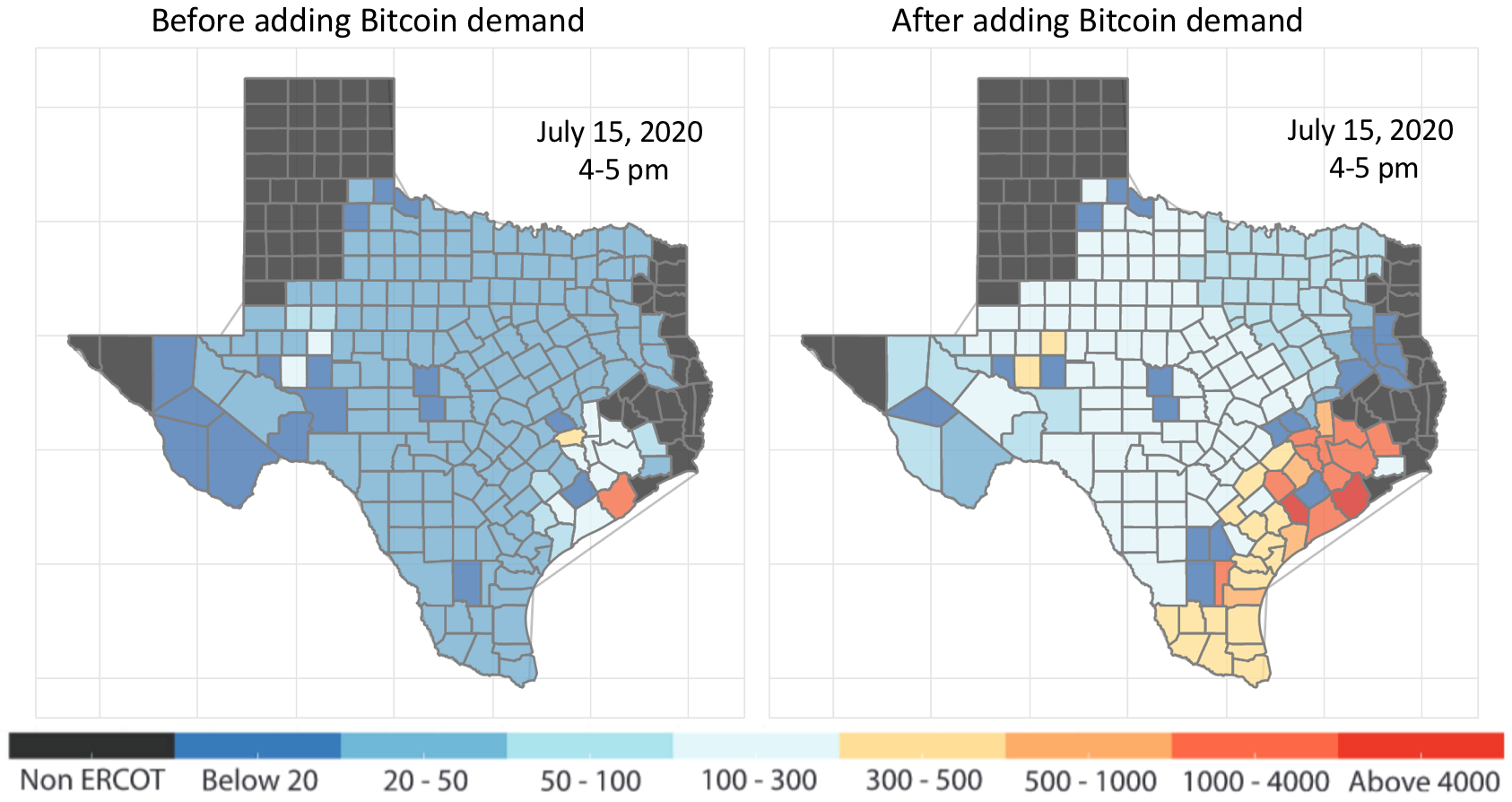}
        \caption[]%
        {{\small 960MW mining loads in Location B}}
        \label{fig:map2}
    \end{subfigure}
    \caption[]
    {\small County-level LMP (\$/MWh) before and after adding mining loads.}
    \label{fig:map}
\end{figure*}

\subsection{Impacts of Mining Loads on LMP}
Remark that LMP is a locational (nodal) price signal at each operating time. We investigate two aspects of LMP, namely the average LMP and hourly LMP, defined as follows. At each time, the average LMP is a value that is averaged over 2000-bus, and the average LMP is investigated over a summer period of 61 days (Days 180 - 240). On the other hand, hourly LMP is averaged again over 61 days for each hour.

The average LMP without additional mining loads is shown in Fig. \ref{fig:original_LMP} for Days 180 - 240. Note that there are several hours that the average LMP is above \$100/MWh. Now, market disruptions of mining facilities are considered. In Location A, there are six mining facilities, and different total capacities 360MW, 480MW, 600MW, 720MW, 840MW are tested out. The average LMP with total capacities 600MW, 720MW, 840MW of mining loads in Location A is shown in Fig. \ref{fig:LMP_locA_600} - \ref{fig:LMP_locA_840} respectively. Observe the peaks of graphs in Fig. \ref{fig:LMP_locA_600} - \ref{fig:LMP_locA_840} are significantly higher than the peaks of a graph in Fig. \ref{fig:original_LMP}. The average LMP with total capacities 360MW, 480MW of mining loads in Location A is similar to but still generally higher than the average LMP in Fig. \ref{fig:original_LMP}. These results are not present to save space.

When total mining loads are higher than 720MW in Location A, SCUC problems start to face ``failed to converge`` in numerical software. Although ``failed to converge`` in numerical software does not always imply that SCUC problem is infeasible because it is possible that the solver is unable to find a solution, it is also observed that higher total capacity renders it failed to converge more frequently. When 840MW, 960MW are tested out in Location A, SCUC problems with 840MW failed to converge 18 days while SCUC problems with 960MW failed to converge 42 days, which is a certificate that the system is pushed towards its limits. Thus, we believe SCUC problems are at the limit boundary (close to infeasible operating point) around the total capacity of 720MW in Location A with the load control strategy.

\begin{table*}[ht]
    \centering
    \caption{LMP characteristics of different settings.}
    \begin{tabular}{lccc} \toprule
         & \begin{tabular}[c]{@{}l@{}}Average LMPs (\$/MWh)
\end{tabular} 
         & \begin{tabular}[c]{@{}l@{}}LMPs during 3-5 pm (\$/MWh)\end{tabular} 
          & \begin{tabular}[c]{@{}l@{}}LMP standard deviation \end{tabular} 
          \\ \midrule 
        Baseline (without mining load) & 27.18  & 42.60 &11.63 
 \\
        Location B with 840MW load & 27.95  & 48.04 & 13.42 
 \\
 Location A with 840MW load & 30.19 &  49.10  &51.06
         \\
        \bottomrule
    \end{tabular}
    \label{tab:lmps}
\end{table*}

Second, market disruptions of mining facilities in Location B are considered. There are six mining facilities, and different total capacities 600MW, 720MW, 840MW, 960MW are tested out. The average LMP with total capacities 840MW, 960MW of mining loads in Location B is shown in Fig. \ref{fig:LMP_locB_840} - \ref{fig:LMP_locB_960} respectively. Observe that the average LMP in Fig. \ref{fig:LMP_locB_840} - \ref{fig:LMP_locB_960} is generally higher than the average LMP in Fig. \ref{fig:original_LMP}.

Now, market disruptions of a combination of mining facilities in Location A and B are considered. There are twelve mining facilities in Location A and B and total capacities 960MW, 1080MW, \ldots, 1440MW are tested, i.e., each facility equally consumes 80MW, 90MW, \ldots, 120MW, respectively. Hourly LMP for a combination of Locations A and B is shown in Fig. \ref{fig:hourly_lmp} where market disruptions by mining loads are summarized.

Observe in Fig. \ref{fig:hourly_lmp_locA} that hourly LMP of Baseline (without mining loads) is consistently lower than hourly LMP with mining loads in Location A. It is observed that the higher the total capacity, the higher the hourly LMP, especially during peak hours (3pm - 7pm) that hourly LMP significantly changes during peak hours by mining loads, while the hourly LMP is similar during non-peak hours. Moreover, there is a significant change in hourly LMPs between total capacity 720MW and 840MW in Location A. This is largely because the average LMP could significantly change due to system constraints such as transmission line constraints, which create non-linear changes on LMP. Hourly LMP for mining facilities in Location B shows similar results that the higher the total capacity, the higher the hourly LMP, as shown in Fig. \ref{fig:hourly_lmp_locB}. Hourly LMP for mining facilities in Locations A and B is shown in Fig. \ref{fig:hourly_lmp_locAB} that bears the same message.

We mark the importance of locations in electric power grids when integrating a new load resource. Under the load control strategy, we report that Location C has significantly less hosting capacity that it starts to face ``failed to converge`` for far less total capacity compared to Locations A and B. In Location A, total capacities 360MW, 480MW, 600MW, 720MW, 840MW are tested out, and 0, 0, 0, 0, 6, 18 days ``failed to converge`` out of 61 days, respectively; In Location B, total capacities 600MW, 720MW, 840MW, 960MW are tested out, and 0, 0, 0, 4 days ``failed to converge`` out of 61 days respectively; In Location C, total capacities 50MW, 100MW 150MW are tested out, and 2, 44, 59 days ``failed to converge`` out of 61 days respectively; Without mining loads, 61 days are all feasible. Thus, we conclude that Location C has significantly less hosting capacity limits compared to other locations under the load control strategy. Although it is definitely possible that the hosting capacity limit will increase with more sophisticated load control strategies in Location C, however, it is also possible that other locations would have higher hosting capacity.

Lastly, the impact of integrating new mining loads on electricity market is twofold. It increases average LMPs, while also creating large price fluctuations.
As shown in Table \ref{tab:lmps}, with adding new mining loads in Location A, we observe an average increase of nearly \$3 in electricity prices. However, this average is taken across all hours, so while the average price increase might seem insignificant, the increase in peak hour prices is substantially higher, as shown in Fig. \ref{fig:hourly_lmp}. In particular, in Location A for average LMPs between 3pm and 5pm, we observe an increase of \$6.5, as shown in Table \ref{tab:lmps}. In this table, the standard deviation of hourly LMPs characterizes price fluctuations created by mining loads. It is worth noting that new loads in Location A create significant price fluctuations as compared to baseline and Location B. Hence, the value and necessity of demand response is greater in Location A. 

It is worth noting that although we do not explicitly compare different demand response algorithms, our simulation results compare the existing system without mining loads versus having different amounts of mining loads. For example, in Fig. \ref{fig:hourly_lmp_locB}, the amount of mining load changes between 0 and 840 MW. It can be easily seen that when mining facilities participate in a demand response program, their operational capacity and impact on the grid lie within our possible scenarios.
\subsection{County-wise Locational Marginal Prices}

Previous subsection introduces temporal properties of LMP over a summer period, whereas this subsection introduces spatial properties of LMP on a map of Texas. County-level LMPs are illustrated on a map of Texas to visualize the disruptive impacts of mining loads and spatial properties of LMPs. Each county includes more than one node (bus), so to obtain the county-level LMP, LMPs are averaged over buses in each county.

Two snapshots are taken in the study period that we closely look at, Day 197 and 240. County-level LMPs are shown in Fig. \ref{fig:map1} and \ref{fig:map2} and each figure shows county-level LMPs with and without additional mining loads. It is observed in the figures that there is a significant increase in electricity prices for many counties. Additional 840MW of mining loads in Location A makes significantly high LMPs for many counties at 2-3pm Aug 27, 2020 (Day 240). Similarly, additional 960MW of mining loads in Location B makes significantly high LMPs for many counties at 4-5pm July 15, 2020 (Day 197). Significant changes are due to transmission line physical constraints, which create a non-linear effect on the LMPs across the system. Thus, price volatility induced by mining loads can easily disrupt many other market participants in electricity markets. Mining loads, thus, may hold responsibilities as well as potential for ancillary services, including demand response programs, due to its property that it is non-essential and non-critical and thus, can be interrupted. 


\subsection{Quantifying Demand Response Profit}
This section aims to quantify the profit of participating in demand response for cryptomining facilities. The profit is calculated as the reward from participating in demand response subtracted by the loss of not mining cryptocurrency during deployment. While RRS and ERS resources are deployed during large supply scarcity and grid emergency events, price-driven demand response is designed to keep the average LMPs across the system as low as possible. In this program, if the average LMP goes beyond a certain threshold, the system operator deploys these resources. Cryptomining facilities participating in price-driven demand response are rewarded according to their availability, independent of deployment. Mining facilities prefer larger deployment thresholds because it leads to less frequent deployments, giving them more time to keep mining.

\begin{figure}   

\begin{center}
\includegraphics[width=1\columnwidth]{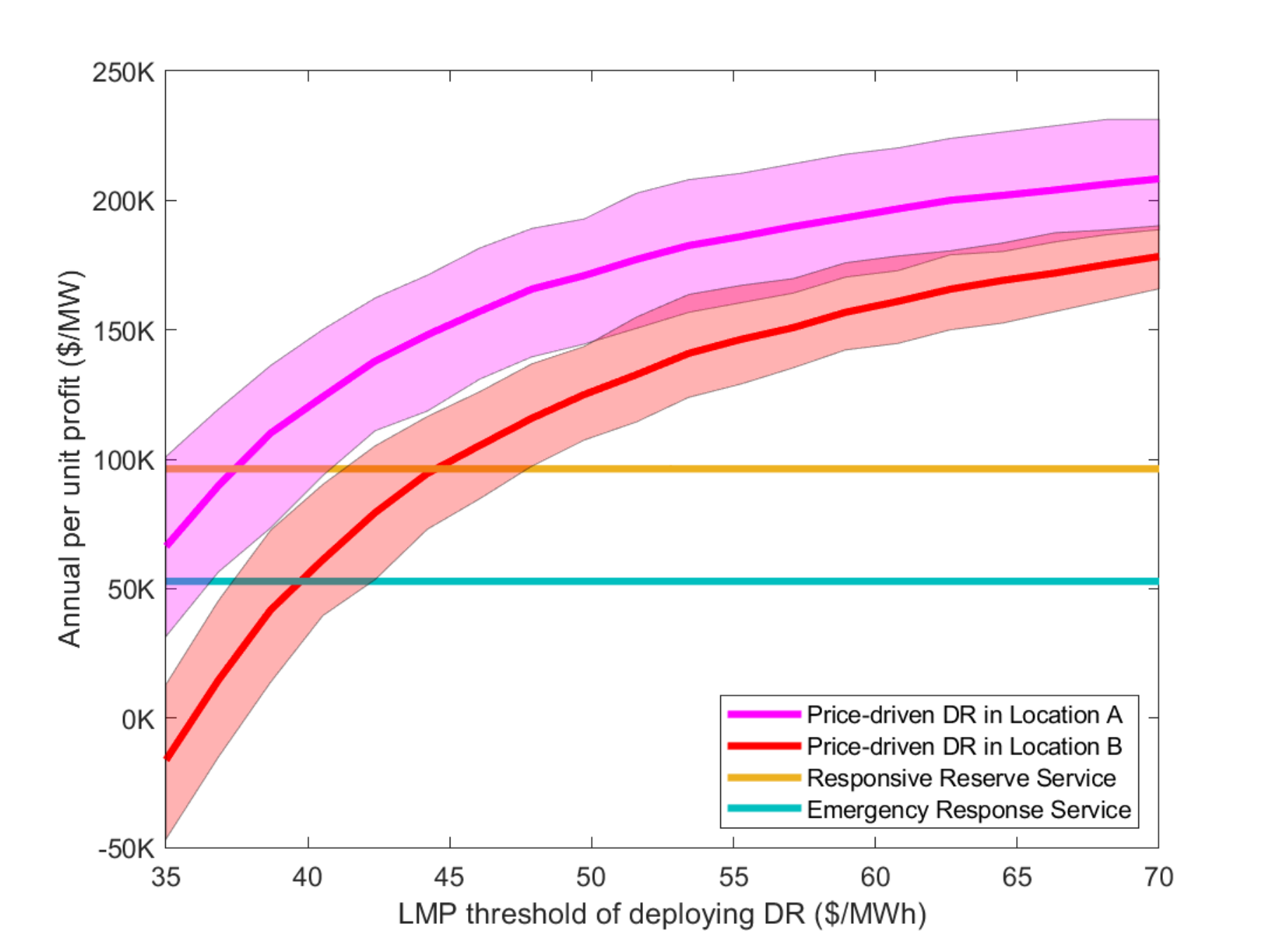}
\par\end{center}
\caption{\label{fig:Profit}Annul per unit profit for one MW of demand response capacity. The x-axis is the price threshold in which the grid operator deploys demand response resources, and the y-axis is the annul per unit profit.}
\vspace{-3mm}
\end{figure}
ERS and RRS deployment records and their corresponding reward data are obtained from ERCOT \cite{Potomac, 2021annual, ercot_dec}, and we assume that the reward of the price-driven demand response is equal to the average hourly LMPs. This is a reasonable assumption because the need for demand response is closely tied with average LMPs in the system, so when electricity prices increase, the demand response rewards also increase. We obtain bitcoin price records from \cite{Bitcoin_price} to calculate the loss of revenue during demand response deployments. Using this information, and by running the synthetic electricity market with 840MW of added mining load in Locations A and B, we calculate hourly LMPs and compare the annual profit of participating in ERS, RRS, and price-driven demand response. In real-world, electricity prices are highly uncertain, and to capture this uncertainty, we obtained ERCOT electricity price records for the year 2019 to 2021 from \cite{LMP_record} and used their distribution to add random noise on top of the hourly LMPs obtained by solving SCED. We repeat our simulations 1000 times and calculate the mean, upper bound, and lower bound of the demand response profit.

In Fig. \ref{fig:Profit} the annul per unit profit for one MW of demand response capacity is plotted. The x-axis is the LMP threshold in which the demand response resources would be deployed, and the y-axis is the per-unit profit. The deployment of ERS and RRS is independent of the threshold, so they get constant earnings of nearly \$50k and \$100k per year, respectively. For price-driven demand response, the larger the threshold, the less deployment there is, which means the miners keep mining, and they get more profit. The upper and lower bounds of the price-driven demand response are also shown in Fig. \ref{fig:Profit}, where their difference comes from added noise on top of the hourly LMPs and shows the impact of price uncertainty on the annual profit. It is worth noting that in Fig. \ref{fig:Profit} the annual profit of price-driven demand response for cryptominers in Location A is more than Location B. This is because, for the same amount of cryptomining loads, the increase in electricity prices in Location A is more substantial than B. Hence, performing demand response in Location A is more valuable, as suggested by our results in Table \ref{tab:lmps}. It can be seen that ERS and RRS are risk-averse programs, while price-driven demand response has a more dynamic structure. If the deployment threshold of the program is too small, resources would be deployed frequently, which might lead to negative profits, as shown in Fig. \ref{fig:Profit}.

%% file: 7-Discussion.tex
\section{Discussion}
\label{sec:Discussion}
In this section, we summarize our findings and discuss the possible implication of our results.
With added cryptomining loads comes the possibility of higher LMPs and more price fluctuations. The impact of these loads on LMPs is inherently different depending on the location of the loads. By careful reduction of cryptomining loads during certain hours in specific locations, one could potentially mitigate these adverse impacts. There are public expectations for cryptominers to actively participate in demand response programs, and the numerical results show that it could be a win-win strategy for both cryptominers and the system operator. Miners could create a substantial revenue stream by providing flexibility, which in turn reduces electricity prices across the system.

Integration of new mining loads into any power grid requires extensive interconnection analysis and infrastructure planning. This type of planning goes beyond grid operators and energy sector. The policymakers face the dual challenge of keeping up with the fast pace of the cryptomines moving into Texas while ensuring that the infrastructure is ready and the grid security is not compromised. Without proper planning, these loads could increase electricity prices, impact voltage and frequency control in the system, and affect grid reliability, particularly during summer peak hours. Currently, there are abundant wind energy resources in West Texas, and co-locating mining loads with these resources could potentially help the grid by reducing renewable curtailment and avoiding major investments in developing new transmission lines connecting West Texas to other regions. Our results suggest that because of the decisive impact of location, it is imperative to design proper incentives encouraging new mining facilities to be built in places with greater societal benefits. Having mining facilities invest in renewable generation units to cover a certain percentage of their operational capacity could also help reduce their carbon footprint, and locational marginal prices.

%% file: 9-Conclusion.tex
\section{Conclusion and Future Work}
\label{sec:conc}

This paper investigated the impact of integrating new cryptocurrency mining loads on a synthetic Texas power grid. We formulated and analyzed how this demand flexibility could improve grid reliability by participating in demand response. The importance of large load interconnection studies to ensure reliable and optimal grid operation was discussed. Our results suggest that the impact of new mining loads on increasing price fluctuations is highly non-uniform. We also quantified the potential profit of participating in different demand response programs for mining facilities. Future work will examine the impact of cryptocurrency miners co-located with other sources in the energy system. Another interesting future direction is to design proper incentive mechanisms in electricity markets to maximize the value and participation of mining loads in provision of demand flexibility. 